\documentclass[runningheads]{llncs}
\usepackage{graphicx}
\usepackage{comment}
\usepackage{amssymb,amsmath}
\usepackage{listings}
\usepackage{color}
\usepackage[most]{tcolorbox}

\lstdefinestyle{MyAsmStyle} {language={[x86masm]Assembler},	morekeywords={JMPZ, JMPN, JMP, LOAD, STORE, DIV, MULT, SUB, ADD, HALT}}
\begin{document}
\title{Verifiable Computations with RAM-like Running Times}

%
%
\author{Tahsin C. M. D{\"o}nmez
	\orcidID{0000-0003-4881-3206}
}
\authorrunning{T. C. M. D{\"o}nmez}
%
\institute{Department of Future Technologies, University of Turku, Turku, Finland
\email{tcmdon@utu.fi}}
\maketitle              
\begin{abstract}
Current and emerging trends such as cloud computing, fog computing, and more recently, multi-access edge computing (MEC) increase the interest in finding solutions to the verifiable computation problem. Furthermore, the number of computationally weak devices have increased drastically in recent years due to the ongoing realization of the Internet of Things. This work proposes a solution 
which enjoys the following two desirable properties: (1) cost of input preparation and verification is very low (low enough to allow verifiable outsourcing of computations by resource-constrained devices on constrained networks); (2) the running time of the verifiable computation is RAM-like.
\end{abstract}

\section{Introduction}

Verifiable outsourcing of computations involves a possibly computationally weak outsourcing party (outsourcer), and one or more worker parties (evaluators) who are possibly untrusted by the outsourcer. The outsourcer sends the inputs for the computation to the evaluator, and the evaluator sends back the result of the computation along with some additional information which enables the outsourcer to verify the received result. How much the outsourcer benefits from outsourcing depends on how much less the cost of verification is compared to the cost of performing the computation, $Cost_{C}$. Obviously, if the cost of verification is greater than or equal to $Cost_{C}$, the outsourcer would rather perform the computation itself. It is also desirable that, the cost of the verifiable computation to the evaluator is as close as possible to $Cost_{C}$.

Solutions to the verifiable computation problem based on Yao's Garbled Circuit (GC) construction enjoy the non-interactivity and inherent verifiability of secure 2-party computations using GCs, but they have to defeat two great challenges before they can be of practical value: the single-use nature of the garbled circuit, and the inflation of size and running time due to the conversion to Boolean circuit. Simply converting a RAM program to a circuit, and then garbling and evaluating it, leads to solutions with circuit-like running times, which is significantly worse compared to the running time of the original RAM program. The solution presented in this work does not 
address the inflation of size, but it does achieve RAM-like running time. 
The verifiable RAM (VRAM) construction which underlies the solution sits somewhere between the simple conversion to circuit and the intricate GRAM constructions~\cite{gram}. The design of VRAM is based on RAM concepts, but unlike GRAM, all the construction work takes place at compile time at a cost similar to circuit construction. The construction underlying the solution is not oblivious, and the solution does not provide privacy. 

The rest of this paper is organized as follows. Section~\ref{bg} provides necessary background information on the random-access machine and garbled circuits. 
Section~\ref{ram} develops the necessary concepts for describing VRAM, and provides an informal description of it. Section~\ref{vramscheme} describes the algorithms which define the VRAM scheme, and Section~\ref{protoc} puts these algorithms together within a protocol, which serves as the formal description of the proposed solution. Section~\ref{concl} concludes the paper and discusses future work.

\section{Background}
\label{bg}

\subsection{Random-Access Machine (RAM)}
\label{bgram}

\begin{figure}
	\includegraphics[width=\textwidth]{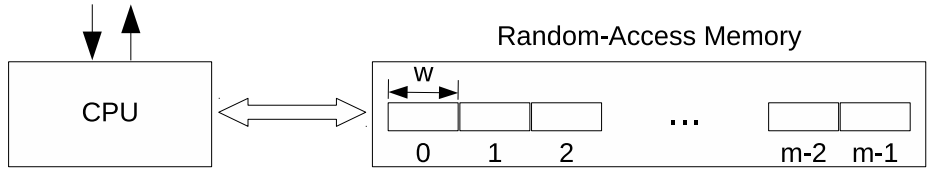}
	\caption{The two main components of the random-access machine: CPU and the random-access memory.} \label{figram}
\end{figure}

The random-access machine (RAM) models the essential features of the traditional serial
computer~\cite{compmodel}. RAM model of computation resembles the operation of modern computers much more closely compared to logic circuits. The random-access machine consists of a central processing unit (CPU) and a random-access memory, which are connected to each other and interact (See Fig.~\ref{figram}). The CPU has a small (compared to the random-access memory) internal memory comprised of special-purpose memory locations called registers, and (for efficiency reasons) all CPU operations are performed on data stored in these registers. The random-access memory is modeled as a collection of $m$ $w$-bit words, each of which is identified by a memory address. The random-access memory stores both data and collections of CPU instructions called programs. The CPU repeatedly reads an instruction from the random-access memory and executes it, modifying data in the process. The set of all instructions comprise the instruction set (IS). A typical IS includes memory load and store instructions for moving data between memory locations and registers, jump instructions, arithmetic and logical instructions, as well as input and output instructions, and a \texttt{HALT} instruction. Branching and loops in high-level languages correspond to \textit{conditional jumps} and \textit{conditional backward jumps}, respectively. 
In a conditional jump, the CPU either reads the next instruction in forward direction, or `jumps' to an instruction out of sequence and reads that one, depending on the result of a comparison. Without loss of generality, the random-access memory can be considered as the union of five disjoint memory regions $R$, $P$, $X$, $Y$ and $D$. The registers will be considered as part of the memory, for the sake of simplifying notation. The read-only $X$ and $P$ are the regions where the input to the program and the program itself are loaded, respectively. $Y$ is the region where the computation result is written at the end of the computation: without loss of generality, and for reasons that will become clear later, we assume that the last thing a program does is to write the computation result into $Y$. Everything else (e.g. local and global variables) is stored in $D$. Then, a RAM computation can be expressed as $Y=P^{D}(X)$, where $P^{D}$ denotes that the program $P$ can read the initial memory contents of $D$, as well as reading from locations in $D$ having written to those locations itself. While the latter class of actions treat $D$ as merely temporary storage, the ability of the programs to read the initial contents of $D$ qualifies it as \textit{persistent memory} which persists between executions of several possibly different programs. $R$ on the other hand is not persistent, and a program $P$ should read a location in $R$ only if it has written to it.

Compared to the most efficient equivalent circuit, a RAM program has significantly better average running time, as the circuit evaluation involves (1) evaluating both branches for each branching, and (2) running each loop the maximum possible number of times it can run. On the other hand, RAM execution is not oblivious (to the inputs) while the circuit evaluation is. However, while obliviousness is a desirable property for private computations, that is not necessarily the case for verifiable computations.

\subsection{Garbled Circuits (GC)}

Yao's Garbled Circuit (GC) construction~\cite{yao,yao2} has given rise to numerous research papers, mostly in the area of secure computation. In later years, the original idea has been formalized under the name garbling schemes~\cite{garbscheme}. A garbling scheme comprises five algorithms $Gb,En,Ev,De,ev$ such that: (1) $(F,e,d) \leftarrow Gb(1^{k}, f)$, where $Gb$ is given a security parameter, and the function $f$ which is to be computed. $Gb$ yields $F$, $e$, and $d$ which describe the garbled function, the encoding function, and the decoding function, respectively. (2) $X = En(e, x)$, where $x$ is the input, and $X$ is the garbled input. (3) $Y = Ev(F, X)$, where $Y$ is the garbled output. (4) $y = De(d, Y)$, where $y$ is the un-garbled output. (5) $y = ev(f, x)$. 

\begin{figure}
	\includegraphics[width=\textwidth]{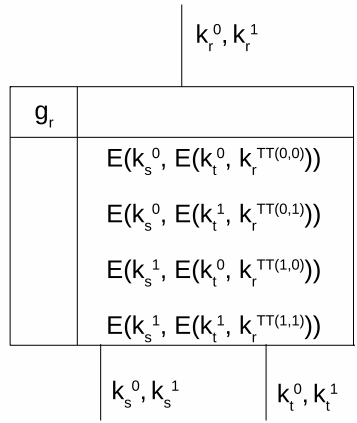}
	\caption{An encrypted gate.} \label{figgate}
\end{figure}

Yao's original construction can be described using the syntax above for garbling schemes as follows. In case of Yao's original construction, $Gb$ garbles a circuit representing a function $f$, and $ev$ is the usual circuit evaluation. $Gb$ starts by assigning to each wire in the circuit, two keys $k^{0}$ and $k^{1}$ corresponding to the two possible wire values $0$ and $1$, respectively. For each gate $g_r$, the keys $k_s^{0}$, $k_s^{1}$, $k_t^{0}$, and $k_t^{1}$ are used to double-encrypt the keys $k_r^{0}$ and $k_r^{1}$, where each one of $s$ and $t$ is either a gate index for a gate whose output wire is connected to an input wire of $g_r$, or an index of an input wire for the circuit. The two encryption keys and the key to be encrypted are chosen respecting the structure of the truth table (TT), so that the evaluation of the garbled circuit with the garbled inputs mimics the in-the-clear evaluation with the corresponding non-garbled inputs. This step is closely related to the correctness condition for the garbling schemes: $De(d, Ev(F, En(e, x))) = ev(f, x)$. The process yields the encrypted truth table (ETT) for the gate (See Fig.~\ref{figgate}). Finally, the rows of the ETTs are shuffled, so that the values on a gate's input wires cannot be inferred from the index of the row opened during the evaluation. In this case, $Ev$ resembles the usual circuit evaluation in terms of the processing order of the gates, however gate evaluations involve undoing the double encryptions, rather than doing simple look-ups in the TTs. Selection of the row to be decrypted may be carried out via trial and error (possible if authenticated encryption is used), or via the point-and-permute technique~\cite{BMR90}. $En$ and $De$ are as simple as following the mappings between the bit values $0$ and $1$, and the corresponding key values.

\section{The Verifiable RAM (VRAM)}
\label{ram}

This section describes the verifiable RAM (VRAM). In order to achieve this, the necessary concepts for describing VRAM will be introduced, and examples will be given for providing context.

A VRAM construction allows one-time verifiable computation. 
It is built on the principle that the execution of a VRAM program can advance so long as the memory access pattern of the VRAM program `mimics' the memory access pattern of the (non-verifiable) RAM program from which the VRAM program was built. Otherwise, the execution shall not advance. The memory access pattern involves not just the locations accessed, but also the type of access (read or write) and the value read or written. 

\subsection{VRAM Random-Access Memory}
\label{vramram}

The random-access memory of the VRAM will be referred to as \textit{encoded memory}, and will be denoted by $M_{v}$. $M_{v}$ holds encodings of the bits of data manipulated by the program, but not the program itself. Using the notation from Section~\ref{bgram}, only the memory regions $R$, $X$, $Y$, and $D$ are encoded, and the VRAM program is never loaded into the encoded memory. $R_{v}$, $X_{v}$, $Y_{v}$, and $D_{v}$ will denote the encoded twins of $R$, $X$, $Y$, and $D$, respectively. In order to keep a one-to-one correspondence between the regions and simplify the descriptions, region $P$ of $M$ will be omitted in the rest of this work:
\begin{align*}
M &= (R || X || Y || D)\\ 
M_{v} &= (R_{v} || X_{v} || Y_{v} || D_{v})
\end{align*}
In case of persistent memory, $X$ and $D$ may both be read as input, affecting the path of execution. The reason for defining a separate region $X$ becomes clear in the context of verifiable computations. It is the region $X$ that stores the inputs of the outsourcing party, whereas $D$ might be some large database whose contents may have been altered by previous computations, and may affect the outcome of the current computation, just as contents of $X$ does. 

If the word length of RAM memory $M$ is $W$, then the word length of $M_{v}$ is $W \cdot K$, where $K$ is the key length, which is the sole security parameter for the VRAM construction. Locations in $M$ and $M_{v}$ are denoted by $x$ and $x_{v}$, respectively. Each bit value stored at location $x$, maps to a key whose first bit is stored at location $x_{v} = x \cdot K$ of $M_{v}$. This mapping from bit values to keys is time-dependent. Time dependency of the mapping is a must because a RAM program may write the same value to a location at different times during execution, but the verifiable twin VRAM relies on garbled circuits for its verifiability property, and garbled circuits require fresh un-exposed keys as inputs. A time-like variable \textit{VRAM time}, denoted by $t$, is incremented by $1$ each time a word in $M$ is written. $t$ also increases due to branchings, as will be explained in the next subsection. 
The last write times $t_{w}[x]$ are separately kept for each memory location $x$, to be used during the construction of the VRAM program. $t_{w}[x] = 0$ for all $x$ at $t=0$, and when some memory location $x'$ is written to at $t=t'$, $t_{w}[x']$ is set to $t'$, whether or not the old and new bit values are different. 

The crucial feature of the encoded memory to keep in mind is that memory writes to $M$ are reflected in the VRAM as \textit{time-translation} of keys, which take place even when the value in $M$ remains unchanged.  

\subsection{VRAM CPU and VRAM Programs}
\label{vramcpu}

It was mentioned in Section~\ref{bgram} that a RAM computation can be expressed as $Y = P^{D}(X)$. Our goal is to obtain a verifiable version of the same computation, which yields $Y_{v} = P_{v}^{D_{v}}(X_{v})$. Previous subsection described how memory is encoded. This subsection describes how the VRAM program $P_{v}$ can be built from $P$. Definition of a separate entity VRAM CPU is not necessary, as the VRAM program will cover the functionality associated with both the CPU and the RAM program $P$. 

A VRAM program consists of several garbled circuits, each belonging to one of the three categories $B$, $T$, or $I$. Type $B$ (branch) and type $T$ (time-merge) circuits together model a conditional jump, and type $I$ (instruction) circuits model any instruction which alters memory. Type $B$ circuits guarantee that only a single branch -the correct one for the given inputs- can be followed, and type $T$ circuits are needed for merging branches, and more generally, for handling input-dependent program behaviour. Type $I$ circuits may be further categorized into sub-types which closely resemble the operations in instruction sets such as x86 and x86\_64, and they guarantee that $M_{v}$ is altered in a way that is consistent with its twin $M$ at each time step, i.e. the memory access pattern is mimicked. Before going any further, we define a few concepts which are relevant to both RAM programs and VRAM programs:
\begin{description}
\item[Segment:] A \textit{segment} is an ordered, maximal-length sequence of instructions which are always executed in sequence, independent of initial memory contents. The sequence order reflects the order in which the instructions are executed. 
\item[Branch:] Either a conditional jump or an \texttt{HALT} instruction marks the end of a segment. In case of a conditional jump, two new segments $s_{1}$ and $s_{2}$ are created, such that at least one of them has non-zero length. The created segments are called \textit{branches}. Let the VRAM times associated with the first and the last instructions in either $s_{1}$ or $s_{2}$ be $t_{1}$ and $t_{2}$, respectively. $t_{1}-1$ (resp. $t_{2}+1$) is defined as the \textit{time of split} (resp. \textit{time of merge}), and is denoted by $t_{s}$ (resp. $t_{m}$). 
\item[Path of Execution:] A \textit{path of execution}, or an \textit{execution path}, is an ordered sequence of segments visited during a single program execution. The sequence order reflects the order in which the segments are visited.
\end{description}
The VRAM time runs from $t=0$ to $t=\tau$ during an execution, where $\tau$ is an input-independent value. Clearly, the input-independent $\tau$ is not a measure of the running time of the RAM, or the VRAM. We define another variable $t_{cost}$, which is more relevant for running time measurements, and use it for imposing a limit on the size of the VRAM program.

\begin{figure}
	\includegraphics[width=\textwidth]{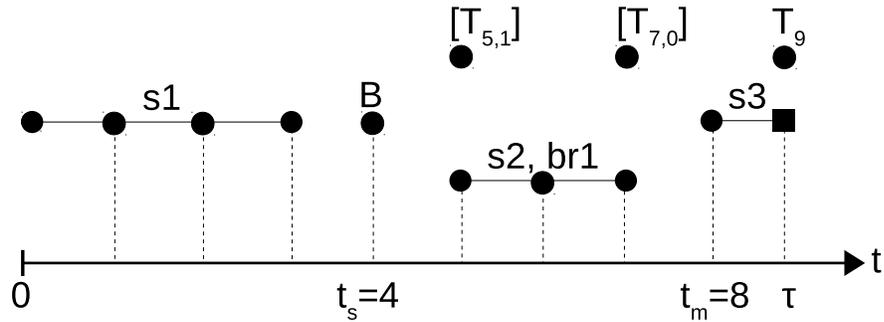}
	\caption{The VRAM program $P_{v}$, corresponding to the RAM program $P$ given in Listing~\ref{codeex1asm}.} \label{figexample1}
\end{figure}

The following example aims to clarify these definitions. First, part of a program $P$ written in an assembly language is given (See Listing~\ref{codeex1asm}). Equivalent code written in a high-level language is given in Listing~\ref{codeex1high}. Finally, the VRAM program $P_{v}$ built from $P$ is depicted in Fig.~\ref{figexample1}.
\begin{lstlisting}[caption={Part of a program $P$ written in an assembly language.},label={codeex1asm},language={[x86masm]Assembler}, style=MyAsmStyle]
;...
LOAD y
STORE x
LOAD z
JMPZ End
LOAD x ;branch 1
ADD one ;branch 1
STORE x ;branch 1
End:
LOAD x
HALT
\end{lstlisting}
\begin{lstlisting}[caption={Equivalent code written in a high-level language.},label={codeex1high},language={C++}]
//...
x = y; //segment 1
if (z != 0) {
x++; //segment 2, branch 1
}
//branch 2 (0-length branch)
return x; //segment 3
\end{lstlisting}
In Fig.~\ref{figexample1}, the axis below shows the VRAM time $t$. Dots indicate garbled circuits. Type $B$ and $T$ circuits are marked with the respective letter, and all unmarked dots correspond to type $I$ circuits. A \texttt{HALT} instruction is marked with a square. In case of type $T$ circuits a single dot is used to represent possibly several $T$ circuits. In other cases, a single dot represents a single circuit. Segments are denoted by $s$, and branches are denoted by $br$. Branch $br1$ and the $0$-length branch $br2$ (which contains only $T$ circuit(s)) both start at $t=5$. Time of split is $t=4$ and time of merge is $t=8$. The square brackets around the $T$ circuits are included to emphasize the fact that existence of $T$ circuits at the end of branches depends on the instructions in both branches. A VRAM time value and a branch index together define a unique circuit position within the structure of a VRAM program. 
We adapt the convention that, $I_{t,b}$ stands for the type $I$ circuit associated with VRAM time $t$, and the upper (resp. lower) branch if $b=0$ (resp. $b=1$). If a circuit is not associated with any branches, $b$ is omitted. Same convention is used also for type $B$ and type $T$ circuits. $I$ and $B$ circuits on the same branch, as well as those that do not belong in any branches, are drawn at the same height. All $T$ circuits are depicted on a vertical line of their own.

A challenge in building a VRAM program is the input dependency of the execution path. Consider the garbled circuit $I_{8}$ in Fig.~\ref{figexample1}. $I_{8}$ takes as input the encoding keys associated with $(\tilde{x},t_{w}[\tilde{x}])$ for all locations $\tilde{x}$ in which bits of the program variable $x$ are stored. These input keys have to be known at compile time so the circuit $I_{8}$ can be constructed. 
The variable $x$ is written at $t=2$, and then at $t=7$ in only one of the branches, which would mean $t_{w}[\tilde{x}]$, and consequently the input keys, depend on the path of execution, which is unknown at compile time. But this is not the case. While building the VRAM program, we make sure that $t_{w}[\tilde{x}]$ is input independent, by \textit{fast-forwarding} keys. Recall that memory writes are modeled by time translation of keys. Fast-forwarding is time translation of keys in order to compensate for time discrepancies due to branching, apart from the normal time translations due to memory writes. Fast-forwards happen in two ways: (1) explicitly via $T$ circuits; (2) implicitly in certain $I$ circuits. $T$ circuit(s) are added to the very end of a branch $br$ when there are memory location(s) $\tilde{x}$ that are modified in the other branch, but not in $br$, explicitly fast-forwarding all $\tilde{x}$ to the time of merge. The implicit case occurs when a memory location $\tilde{x}$ is written by one or more $I$ circuits on a branch. The very last time some $\tilde{x}$ is written on a branch, the $I$ circuit which does the writing does not use the VRAM time associated with it to determine the output keys, but instead uses the time of merge, possibly fast forwarding $\tilde{x}$. There is one other case where explicit fast-forwards occur. A \texttt{HALT} instruction does not alter RAM memory, so it has no corresponding $I$ circuit. It is represented in the VRAM program merely with a marker. These marked positions indicate the end of each possible path of execution (with possibly different running times) at compile time, and program termination at runtime. At these positions are $T$ circuits which fast-forward the whole $Y_{v}$ region to $t=\tau$, making possible the verification of computation result using a single key pair per location. 


While building the VRAM program, we have to ensure that the computing party can follow only the correct path of execution while executing the VRAM program. 
This is achieved by replacing each conditional jump in the RAM program with a circuit which evaluates the condition (e.g. `is zero?' for \texttt{JMPZ}), and outputs one of the two branch keys depending on the result. Each garbled circuit on a branch, regardless of its type, is encrypted with the corresponding branch key. 
Below, we present two more examples before taking a closer look at the garbled circuits involved. In order to save space, we only give the high-level language code. We won't be precise about segment lengths and $t$ values, and will concentrate on the VRAM program structure instead.
\begin{lstlisting}[caption={Example 2. A simple RAM program.},label={codeex2high},language={C++}]
execSegment1();
if (condition1) {
	execSegment2();
} else if (condition2)
	execSegment3(); //last statement is a return
} else {
	execSegment4();
}
execSegment5(); //last statement is a return
\end{lstlisting}
\begin{figure}
	\includegraphics[width=.65\textwidth]{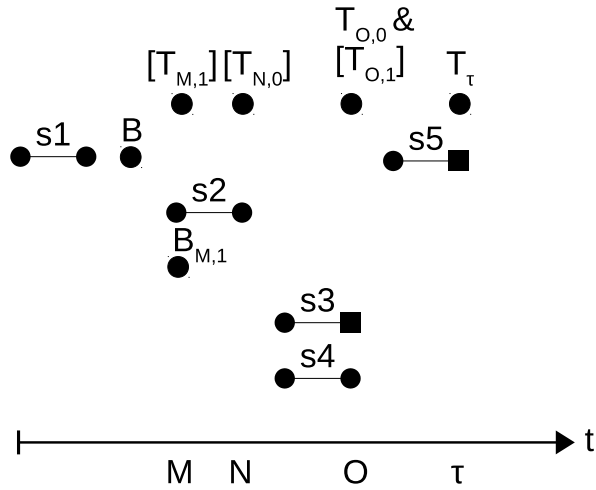}
	\caption{The VRAM program corresponding to the RAM program given in Listing~\ref{codeex2high}.} \label{figexample2}
\end{figure}
Listing~\ref{codeex2high} contains a typical \texttt{if-else} statement. One thing to note in Fig.~\ref{figexample2} is that the first branching is already merged before the second one takes place. In some sense, building a VRAM program involves flattening the associated RAM program into two-branch thickness, by considering the expanded VRAM time instead of the regular running time of a RAM program. In general, a VRAM program handles at most two branches at each VRAM time $t$. Another thing to note in this example is that the RAM program includes two return statements (i.e. \texttt{HALT} instructions), and both $T_{O,0}$ and $T_{\tau}$ fast-forward $Y_{v}$ to $t=\tau$.
\begin{lstlisting}[caption={Example 3. A simple RAM program.},label={codeex3high},language={C++}]
execSegment1();
while (condition1) {
	execSegment2();
}
execSegment3(); //last statement is a return
\end{lstlisting}
\begin{figure}
	\includegraphics[width=\textwidth]{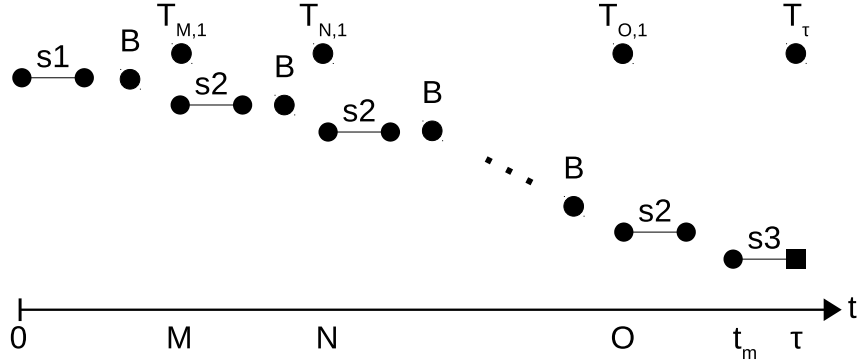}
	\caption{The VRAM program corresponding to the RAM program given in Listing~\ref{codeex3high}.} \label{figexample3}
\end{figure}
Listing~\ref{codeex3high} contains a \texttt{while} loop. The things to note in this example are that: (1) $T_{M,1}$, $T_{N,1}$, ... , $T_{O,1}$ all fast-forward their input keys to the same VRAM time $t=t_{m}$; (2) each $B$ circuit outputs either the branch key to encrypt the immediately following ($s2$ segment, $B$ circuit) pair, or the branch key to encrypt $s3$. 

\subsection{Building the $I$, $B$, $T$ garbled circuits}
\label{circuits}
Circuits of each type have quite simple structure, so we will provide only one example of each. We will assume that (1) $IS = \{\texttt{LOAD}, \texttt{STORE}, \texttt{ADD}, \texttt{SUB}, \texttt{MUL}, \texttt{DIV}, \texttt{JMP},\\
 \texttt{JMPZ}, \texttt{JMPN}, \texttt{HALT}\}$; (2) the instruction \texttt{LOAD} loads its parameter into register $r$; (3) the instruction \texttt{JMPZ} makes the comparison with the value in register $r$. The keys for encoding memory are generated from a pseudo-random function $F^{k}(x,t,b)$, and the branch keys used by $B$ circuits are generated from a PRF $F^{br}(t,br)$. $x$ is a memory location in $M$, $t$ is the VRAM time, $b,br \in \{0,1\}$ are a bit value and a branch index, respectively.\footnote{Note that neither of these PRFs plays the same role it plays in the original GRAM construction, where the PRF key is embedded in some of the circuits to allow part of construction work to be done in runtime.}

First, we construct $I_{3}$ which corresponds to the instruction \texttt{LOAD z} in Listing~\ref{codeex1asm}. $I_{3}$ is an $I$ circuit with sub-type \texttt{LOAD}. Suppose that the location of the word holding variable $z$ is $x_{z}$, and the location of the register $r$ in $R$ is $x_{r}$. In this case, the circuit being built is simply all the circuits for loading individual bits, bundled together.\footnote{This is not always the case. For example, the individual bits mix together due to carry bits for the \texttt{ADD} instruction.} So we consider only the circuit $I^{0}_{3}$ responsible from loading the first bit of $z$ at $x_{z}$. $I^{0}_{3}$ is a gate with a single input wire and a single output wire, whose two ETT rows $R_{0}$, $R_{1}$ are:
\begin{align*}
R_{0} &= E_{k^{0}}(F^{k}_{s_{k}}(x_{r},3,0))\\ 
R_{1} &= E_{k^{1}}(F^{k}_{s_{k}}(x_{r},3,1))
\end{align*}
where $k^{0} = F^{k}_{s_{k}}(x_{z},t_{w}[x_{z}],0)$, and $k^{1} = F^{k}_{s_{k}}(x_{z},t_{w}[x_{z}],1)$.

Next, we consider $B_{4}$, the $B$ circuit which corresponds to the instruction \texttt{JMPZ End} in Listing~\ref{codeex1asm}. $B_{4}$ is a circuit with $W$ input wires, where $W$ is the word length. Two output wires give the branch key (left wire) and branch index (right wire). Let $i^\text{\tiny th}$ bit of $r$ be at location $x^{i}_{r}$. Then the $i^\text{\tiny th}$ input wire of $B_{4}$ accepts as inputs $k_{i}^{0} = F^{k}_{s_{k}}(x^{i}_{r},t_{w}[x^{i}_{r}]=3,0)$ and $k_{i}^{1} = F^{k}_{s_{k}}(x^{i}_{r},t_{w}[x^{i}_{r}]=3,1)$. The left output wire is associated with the keys $k^{up} = F^{br}_{s_{br}}(4,0)$ and $k_{down} = F^{br}_{s_{br}}(4,1)$, where the former is the branch key for the upper, zero-length branch,\footnote{The branch is zero-length, but $k^{up}$ is not useless, as it is used for encrypting $T_{5,1}$.} and the latter is the branch key for the lower branch $br1$. $B_{4}$ is built such that it outputs $k^{up}$ (and $0$ for branch index) only for inputs $(k_{0}^{0}, k_{1}^{0}, \dotsc, k_{W-1}^{0})$, and it outputs $k^{down}$ (and $1$ for branch index) only when all inputs are accepted but the case for $k^{up}$ is not true (i.e. $r \neq 0$). In all other cases, $B_{4}$ may output garbage values.

Finally, we consider one of the $T$ circuits depicted in Fig.~\ref{figexample1} as $T_{5,1}$.\footnote{These circuits emulate the writes that happen only in the other branch. There are $3$ $I$ circuits on the other branch $br1$ which write to at most $3$ different locations, so up to $3$ $T$ circuits may be needed on the other branch.} A $T$ circuit is almost identical to an $I$ circuit with sub-type \texttt{LOAD}, except that the memory location read and written are the same, so the $T$ circuit merely time-translates a single word. Again, a circuit which operates on a word can be thought of as $W$ circuits, each time-translating a single bit, bundled together. The ETT rows $R_{0}$, $R_{1}$ of $T^{i}_{5,1}$ which time-translates the $i^\text{\tiny th}$ bit are:
\begin{align*}
R_{0} &= E_{k^{0}}(F^{k}_{s_{k}}(x_{i},t_{m}=8,0))\\ 
R_{1} &= E_{k^{1}}(F^{k}_{s_{k}}(x_{i},t_{m}=8,1))
\end{align*}
where $k^{0} = F^{k}_{s_{k}}(x_{i},t_{w}[x_{i}],0)$, and $k^{1} = F^{k}_{s_{k}}(x_{i},t_{w}[x_{i}],1)$. Unlike the circuits considered in the previous examples, $T_{5,1}$ is on a branch. What is added to the VRAM program is not the circuit, but the ciphertext resulting from its encryption using the branch key $k^{up} = F^{br}_{s_{br}}(4,0)$.

\section{The VRAM Scheme}
\label{vramscheme}

The VRAM scheme is comprised of the following four algorithms: $\mathcal{A}_{PROG}$, $\mathcal{A}_{INPUT}$, $\mathcal{A}_{VERIFY}$, and $\mathcal{A}_{EXEC}$. These describe construction of a VRAM program, encoding of inputs, verification of a computation result, and execution of a VRAM program, respectively. 
\begin{tcolorbox}[breakable, enhanced, colback=white, colframe=black]
\begin{center}
	\begin{small}
		$\mathcal{A}_{INIT}(\tau_{prev})$
		\noindent\rule{\textwidth}{1pt}
		\begin{description}%
			\item[Description:] A subroutine in $\mathcal{A}_{PROG}$. Prepares $t$, $t_{start}$, $\tau$, and $t_{w}[x]$ 
			for a new computation. 
			\item[Preconditions:] PRF $F^{k}$, $F^{br}$, and the corresponding PRF keys $s_{k}$, $s_{br}$ are fixed. 
			\item[Inputs:] $\tau_{prev}$ : the termination time ($\tau$) of the last built (via $\mathcal{A}_{PROG}$) VRAM program.\footnote{Several VRAM programs may be built to operate on the persistent memory $D_{v}$.} 
		\end{description}
		\begin{enumerate}
			\item Set $t_{start} = \tau_{prev}$.
			\item Set $\tau = 0$.
			\item If $\tau_{prev}=0$: 
			\begin{itemize}
				\item Set $t = 0$.
				\item Set $t[w] = 0$ for all $x$ in $M$.
			\end{itemize}
		\end{enumerate}
	\end{small}
\end{center}
\end{tcolorbox}
\begin{tcolorbox}[breakable, enhanced, colback=white, colframe=black]
\begin{center}
	\begin{small}
		$P_{v}, t_{start}, \tau \leftarrow \mathcal{A}_{PROG}(P, \tau_{prev})$ 
		\noindent\rule{\textwidth}{1pt}
		\begin{description}
			\item[Description:] Builds a verifiable version of the given RAM program.
			\item[Preconditions:] $P_{v}$ is empty.
			\item[Inputs:] $P$ : RAM program; $\tau_{prev}$ : the VRAM time at which the last built program ends. $0$ if there is no previously built VRAM program.
			\item[Outputs:] $P_{v}$ : VRAM program. A collection of labeled and possibly encrypted garbled circuits; $t_{start}$ : The VRAM time when $P_{v}$ starts; $\tau$ : The VRAM time when $P_{v}$ terminates.
		\end{description}
		\begin{enumerate}
			\item Prepare for a new computation by running $\mathcal{A}_{INIT}(\tau_{prev})$.
			\item Systematically follow every path of execution for the given RAM program $P$. We will assume the same IS from Section~\ref{circuits}. 
			For each path, do:
			\begin{itemize}
				\item Set $t_{cost} = 0$. 
				\item While $t_{cost} < MAX_{cost}$ and no \texttt{HALT} instruction is encountered, do:
				\begin{itemize}
					\item Pick the next instruction $inst$ on the execution path. 
					\item Increment $t_{cost}$ by $1$.\footnote{We make the simplifying assumption that every instruction has the same cost.}
					\item If $inst \in \{\texttt{LOAD}, \texttt{STORE}, \texttt{ADD}, \texttt{SUB}, \texttt{MUL}, \texttt{DIV}\}$, then build an $I$ circuit with the corresponding sub-type, and update $t_{w}$ with current time $t$ for the memory location written. However, if $inst$ is on a branch, and the target location is written for the last time on this branch (implicit fast-forward), then (1) build $I$ s.t. it time-translates to the time of merge instead of current time; (2) update $t_{w}$ with time of merge for the memory location written. If $inst$ is the final instruction on the branch, build $T$ circuits which emulate the writes that happen only in the other branch (explicit fast-forward), and set $t$ to the time of merge. Otherwise, increment $t$ by $1$.
					\item If $inst \in \{\texttt{JMPZ}, \texttt{JMPN}\}$, then build a $B$ circuit. Increment $t$ by $1$.
					\item If $inst$ is an \texttt{HALT} instruction, then build $T$ circuits which fast-forward each key in $Y_{v}$ to $\tau$. 
					This action is delayed until all paths are exhausted, and $\tau$ is fixed.\footnote{One other possibility is to use a predetermined $\tau$ value which is guaranteed to exceed all $t_{max}$. Using a $\tau$ value which is larger than necessary has no drawbacks.} Add a dummy element to $P_{v}$ with type field in the circuit label set to \texttt{HALT}.
					\item For each circuit built during the processing of $inst$, prepare wire labels and a circuit label. Wire labels associate with each wire the memory location read or written.\footnote{An exception is the output wires of a type $B$ circuit.} Circuit label indicates the VRAM time when the circuit was built, type ($I$,$B$,$T$), and also the branch index ($0$ for `upper' branch, $1$ for `lower` branch) if the circuit is on a branch. We adopt the convention that the branch that is followed when the condition evaluates to \texttt{true} is the upper branch. If a circuit is not on a branch, label it and add it to $P_{v}$. Otherwise, encrypt the circuit with the corresponding branch key and add the labeled ciphertext to $P_{v}$. 
				\end{itemize}
				\item Let the largest $t$ value observed on the path be $t_{max}$. If $t_{max} > \tau$, then set $\tau = t_{max}$.
			\end{itemize}			
			\item Set $\tau_{prev} = \tau$.
			\item Return $P_{v}$, $t_{start}$, $\tau$.     
		\end{enumerate}
	\end{small}
\end{center}
\end{tcolorbox}
\begin{tcolorbox}[breakable, enhanced, colback=white, colframe=black]
\begin{center}
	\begin{small}
		$X_{v} \leftarrow \mathcal{A}_{INPUT}(X, F^{k}, s_{k}, t_{start})$
		\noindent\rule{\textwidth}{1pt}
		\begin{description}
			\item[Description:] Input preparation.
		\end{description}
		\begin{enumerate} 
			\item Let $X(x) = b$, where $b \in \{0,1\}$. For each location $x$ in $X$, set the key at the corresponding location $x_{v}$ in $X_{v}$ to $F^{k}_{s_{k}}(x,t_{start},b)$.
			\item Return $X_{v}$.
		\end{enumerate}
	\end{small}
\end{center}
\end{tcolorbox}
\begin{tcolorbox}[breakable, enhanced, colback=white, colframe=black]
\begin{center}
	\begin{small}
		$Y \leftarrow \mathcal{A}_{VERIFY}(Y_{v}, F^{k}, s_{k}, \tau)$
		\noindent\rule{\textwidth}{1pt}
		\begin{description}
			\item[Description:] Verification.
		\end{description}
		\begin{enumerate} 
			\item For each location $y$ in $Y$:
			\begin{itemize}
				\item Let the key at the corresponding location $y_{v}$ in $Y_{v}$ be $k_{y}$. Let $F^{k}_{s_{k}}(y,\tau,b) = k_y^b$, $b \in \{0,1\}$.
				\item If $k_y^b = k_{y}$, set $Y(y) = b$.
				\item If $k_{y} \notin \{k_y^0, k_y^1\}$, then the verification failed. Return $\bot$.
			\end{itemize}
			\item If $k_{y} \in \{k_y^0, k_y^1\}$ for all $y$, then $Y$ is accepted as the verified result of the computation. Return $Y$.
		\end{enumerate}
	\end{small}
\end{center}
\end{tcolorbox}
\begin{tcolorbox}[breakable, enhanced, colback=white, colframe=black]
\begin{center}
	\begin{small}
		$Y_{v} \leftarrow \mathcal{A}_{EXEC}(P_{v}, X_{v}, D_{v})$
		\noindent\rule{\textwidth}{1pt}
		\begin{description}
			\item[Description:] Execute the VRAM program. Execution may alter $D_{v}$, as well as $Y_{v}$.
			\item[Inputs:] $P_{v}$ : VRAM program. A collection of labeled and possibly encrypted garbled circuits; $X_{v}$ : Inputs to the VRAM program; $D_{v}$ : Encoded persistent memory. 
			\item[Outputs:] $Y_{v}$ : Encoded computation result.
		\end{description}
		\begin{enumerate}
			\item Set variable $br=-1$. $br$ holds current branch index at any time during execution. Value $-1$ stands for no branch.
			\item Set variable $k_{br}=null$. $k_{br}$ holds the branch key for the current branch.
			\item Set variable $halt=\texttt{false}$.
			\item Execution involves iterating over and processing the elements in the collection $P_{v}$. Most likely, only a small portion of the elements which are associated with a single path of execution have to be processed. Order of processing is determined by the labels. The element which is picked next for processing is the one with the smallest VRAM time in its label. If $br \neq -1$, only elements with same branch index $br$ 
			in its label can be picked. An element is only picked once. If more than one elements are eligible for picking (i.e. they are labeled with the same VRAM time), type $I$ and $B$ circuits are picked before type $T$ circuits. Elements with type set to $\texttt{HALT}$ are picked last. In other cases, e.g. among types $I$ and $B$, the pick can be made arbitrarily. While there is an element eligible for picking:
			\begin{itemize}
				\item Pick the next element.
				\item If $br \neq -1$, decrypt the circuit using $k_{br}$.
				\item For each input wire, read from $M_{v}$ the key associated with the location in the wire label. Assign each key read to the corresponding input wire. Evaluate the garbled circuit. 
				\item Read circuit type from circuit label.
				\item If type is $I$ or $T$, then for each output wire, read from the wire label the write location, and write the key assigned to the output wire (i.e. the evaluation result) to the corresponding location in $M_{v}$. If type is $T$ and there are no more eligible elements labeled with the same $t$, then set $br=-1$ and $k_{br}=null$.
				\item If type is $B$, then set the key on the left output wire to $k_{br}$, and set the value on the right output wire to $br$. 
				\item If type is $\texttt{HALT}$, then set $halt=\texttt{true}$.
			\end{itemize}
			\item If $halt=\texttt{false}$, return $\bot$. ($MAX_{cost}$ was chosen poorly) 
			\item Return $Y_{v}$.
		\end{enumerate}
	\end{small}
\end{center}
\end{tcolorbox}
Note that extra work has to be done for a branch $br$ that is not executed. The extra work is proportional to the number of distinct memory locations accessed exclusively in $br$, and is independent of the running time of $br$. 
The extra work that has to be done for a loop is proportional to the number of times it is executed. The verifiable RAM program terminates at exactly the same point along the path of execution as its non-verifiable counterpart. The running time of a VRAM program is RAM-like. 

\section{Protocol for Outsourcing VRAM Programs}
\label{protoc}
This section presents a protocol for verifiable outsourcing of computations on persistent memory. The protocol consists of a preprocessing phase and an online phase, and works in a three-party setting. The parties involved are the outsourcer (a possibly computationally weak party who outsources the computations and verifies the results), the evaluator (a computationally capable untrusted party who performs the computations), and the constructor (a computationally capable trusted party who builds the verifiable programs corresponding to the outsourced computations).\footnote{By (un)trusted we mean (un)trusted by the outsourcer.} 
\subsection{Preprocessing Phase}
The \textit{constructor} prepares all the preprocessing material without the involvement of the \textit{outsourcer} and the \textit{evaluator}, who may receive their share of the preprocessing material anytime before the first outsourced computation begins, and possibly at different times.
\begin{tcolorbox}[breakable, enhanced, colback=white, colframe=black]
\begin{center}
	\begin{small}
		Preprocessing Phase
		\noindent\rule{\textwidth}{1pt}
		\begin{enumerate}
			\item $\mathbf{P_{v}, t_{start}, \tau \leftarrow \mathcal{A}_{PROG}(P, \tau_{prev})}$ \textit{Constructor} builds the VRAM program. 
			\item \textit{Constructor} sends $F^{k}$, $s_{k}$, $t_{start}, \tau$ to the \textit{outsourcer}.
			\item \textit{Constructor} sends $P_{v}$, $X_{v}$, $D_{v}$ to the \textit{evaluator}.
		\end{enumerate}
	\end{small}
\end{center}
\end{tcolorbox}

\subsection{Online Phase}
Parties involved in the online phase are the \textit{outsourcer} and the \textit{evaluator}.
\begin{tcolorbox}[breakable, enhanced, colback=white, colframe=black]
\begin{center}
	\begin{small}
		Online Phase
		\noindent\rule{\textwidth}{1pt}
		\begin{enumerate}
			\item $\mathbf{X_{v} \leftarrow \mathcal{A}_{INPUT}(X, F^{k}, s_{k}, t_{start})}$ When \textit{outsourcer} wants to outsource a computation, it decides on its inputs $X$, prepares inputs for the VRAM program, and initiates the computation by sending $X_{v}$ to \textit{evaluator}.
			\item $\mathbf{Y_{v} \leftarrow \mathcal{A}_{EXEC}(P_{v}, X_{v}, D_{v})}$ \textit{Evaluator} executes the VRAM program and sends the result $Y_{v}$ to \textit{outsourcer}.
			\item $\mathbf{Y \leftarrow \mathcal{A}_{VERIFY}(Y_{v}, F^{k}, s_{k}, \tau)}$ \textit{Outsourcer} verifies the received result. If $\mathcal{A}_{VERIFY}$ returns $\bot$, \textit{outsourcer} concludes that \textit{evaluator} tried to cheat, and the protocol terminates. Otherwise, \textit{outsourcer} accepts $Y$ as the verified result of the computation.
		\end{enumerate}
	\end{small}
\end{center}
\end{tcolorbox}

\section{Conclusion and Future Work}
\label{concl}
This work proposed a solution to the verifiable computation problem which accepts resource-constrained devices as outsourcers, and offers RAM-like running times to evaluators. The other side of the coin is that the computational and memory costs of building VRAM programs incurred on the constructor, the cost incurred on the network due to the size of the VRAM programs, or the memory cost of storing VRAM programs incurred on the evaluator might not be tolerable. Moreover, a VRAM program can be used only once. However, there is also reason to be hopeful. First of all, the possibly intolerable costs mentioned above all concern the preprocessing phase of the protocol, and the online phase of the protocol is efficient. Secondly, it seems possible that the memory and communication costs associated with the constructor and evaluator can be made amortizable over several computations.



%
%
%
\bibliographystyle{splncs04}
\bibliography{samplepaper}

\end{document}